\documentclass{PoS}
\usepackage{amsmath,amstext}
\usepackage{lineno}

\title{Limits on the Emission of Gamma Rays from M31 (The Andromeda Galaxy) with HAWC}

\ShortTitle{Search for M31 ``Fermi Bubbles'' with HAWC}

\author{Ryan Rubenzahl$^a$, Segev BenZvi$^a$, and \speaker{Joshua Wood}$^b$ for the HAWC Collaboration$^c$\\
        $^a$Department of Physics and Astronomy, University of Rochester, Rochester, New York, USA\\
        $^b$Department of Physics, University of Wisconsin, Madison, Wisconsin, USA\\
        $^c$For a complete author list, see \href{http://www.hawc-observatory.org/collaboration/icrc2017.php}{http://www.hawc-observatory.org/collaboration/icrc2017.php}\\
        E-mail: \email{rrubenza@u.rochester.edu}, \email{sybenzvi@pas.rochester.edu}, \email{jwood@icecube.wisc.edu}}

\abstract{The detection of the Fermi Bubbles suggests that spiral galaxies such as the Milky Way can undergo active periods. Using gamma-ray observations, we can investigate the possibility that such structures are present in other nearby galaxies. We have analyzed the region around the Andromeda Galaxy (Messier Catalog M31) for signs of bubble-like emission using TeV gamma-ray data recorded by the High-Altitude Water Cherenkov Observatory. We fit a model consisting of two 6 kpc bubbles symmetric about and perpendicular to the M31 galactic plane and assume a power-law distribution for the gamma-ray flux. We compare the emission from these bubble regions to that expected from structures similar to the Fermi Bubbles found in the Milky Way. No significant emission was observed. We report upper limits on the TeV flux from Fermi Bubble structures in M31.}

\FullConference{35th International Cosmic Ray Conference – ICRC217-\\
		10-20 July, 2017\\
		Bexco, Busan, Korea}

\begin{document}

\section{Introduction}
The ``Fermi Bubbles'' are two large spherical regions of gamma-ray emission
oriented symmetrically above and below the Galactic Plane of the Milky Way
\cite{Su:2010qj}. The gamma-ray flux from the bubbles has been observed between 0.1 and 300 GeV and is best characterized by a power law spectrum with a cutoff above 100~GeV \cite{Fermi-LAT:2014sfa}. Their discovery was one of the most unexpected and puzzling gamma-ray observations of the last decade.  The origin of the bubbles and the processes powering their energy output are not understood, although they are likely to be reminiscent of past activity of the supermassive black hole located at the Galactic Center \cite{Su:2010qj}.

At present, only the Milky Way is known to contain very extended regions of
gamma-ray emission like the Fermi Bubbles. However, since most spiral galaxies are believed to contain supermassive black holes at their centers, they may also undergo periods of high activity that give rise to bubble structures. An obvious search target for gamma-ray bubbles is Andromeda (M31), our nearest neighboring large spiral galaxy. A recent analysis based on Fermi-LAT data \cite{Pshirkov:2016qhu} indicates that M31 does contain two $6-7.5$~kpc circular regions of gamma-ray emission between $0.1-300$~GeV. Like the Fermi Bubbles in our own Galaxy, these regions are oriented perpendicular to the galactic plane of M31 and are symmetric about the galactic nucleus.

At TeV energies, the VERITAS Collaboration observed the region around M31 for 54 hours and searched for evidence of emission from the galactic plane and from bubble-like regions around the galactic nucleus \cite{Bird:2015npa}. No significant emission from M31 was observed. However, due to its proximity M31 is a very spatially extended source ($3.2^\circ \times 1^\circ$ \cite{Bird:2015npa}) and this increases the difficulty of the background subtraction procedure used by VERITAS. However, the High-Altitude Water Cherenkov (HAWC) Observatory is well-suited to perform unbiased measurements of gamma rays from very spatially extended objects. HAWC is a continuous survey of the Northern Hemisphere in the 0.1-100 TeV range that consists of an array of 300 water Cherenkov detectors that observe gamma-rays using the water-Cherenkov method. Moreover, M31 is located at $(\alpha_{J2000} = 0^h 43^m 35^s.43, \delta_{J2000} = +41^\circ 20' 56.8'')$, transiting $22^\circ$ from zenith in the HAWC detector. Thus we can use observations from HAWC to look for signs of TeV emission from bubble-like regions.

\section{Extrapolation of Milky Way Bubbles}
\begin{figure}[t]
\centering
\includegraphics[width=\textwidth]{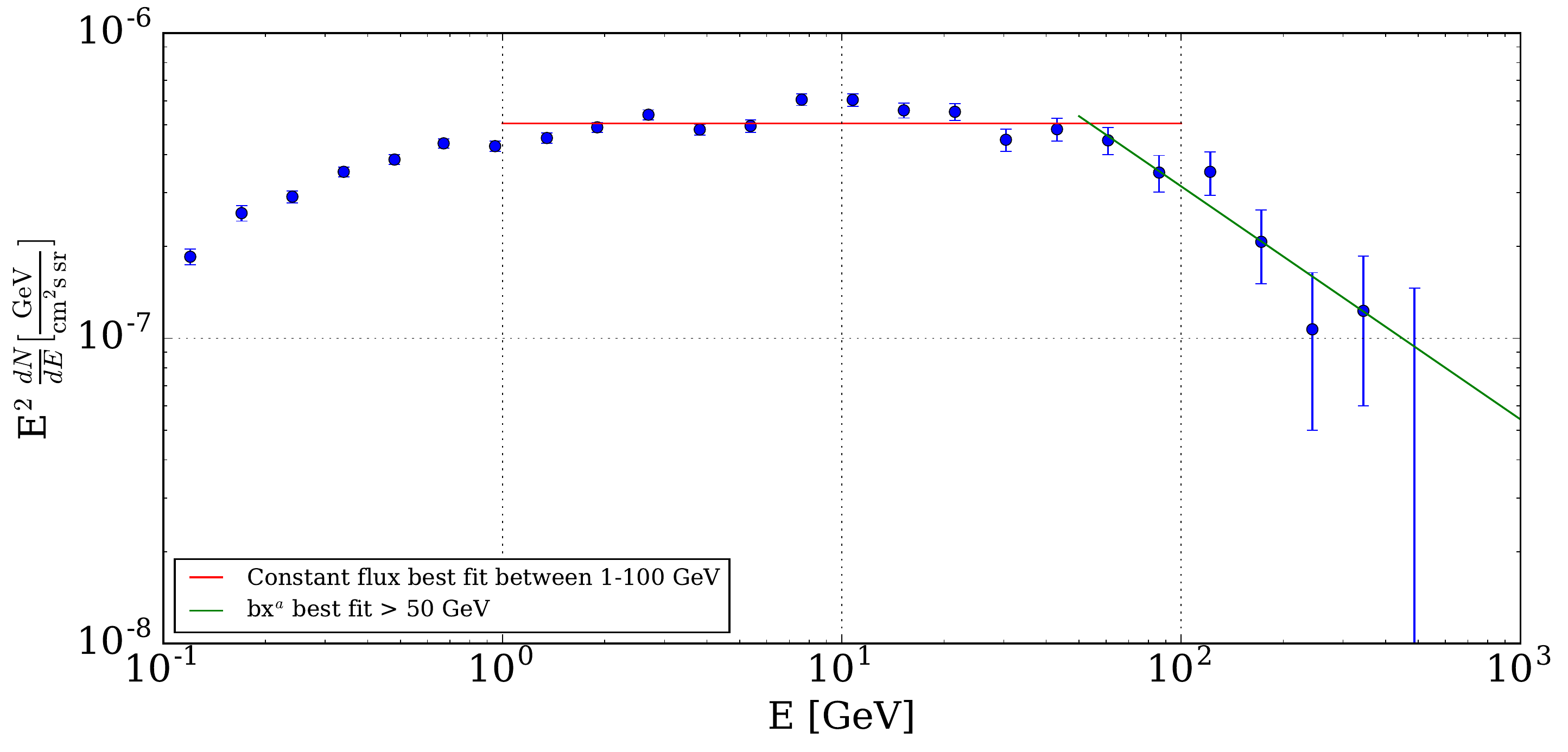}
\caption{Spectral energy distribution (SED) of the Milky Way Fermi Bubbles. Data from Table 2 in \cite{Fermi-LAT:2014sfa}. The red line is the best-fit constant value between 1-100 GeV, corresponding to a spectral index of $-2.0$. The green line is a power law fit to the data above 50 GeV, corresponding to a spectral index of $-2.75$.}
\label{fig:SED}
\end{figure}

The Fermi Bubbles were first detected in a search for residual gamma rays between 1 and 100~GeV measured by the Fermi-LAT detector \cite{Su:2010qj}. The bubbles subtend approximately 0.8~sr above and below the Galactic Center and are located $r_\text{MW} = 9.6$ kpc from the Sun. In the 1~-~100~GeV energy range, the spectral energy distribution (SED) is approximately constant. However at higher energies the spectrum falls off as seen in Figure~\ref{fig:SED} (reproduced from \cite{Fermi-LAT:2014sfa}).

From the observed flux for the Milky Way bubbles, we may calculate the gamma-ray flux of an equivalent bubble-like structure from M31 given the distance $r_\text{M31} =780$~kpc between M31 and the Milky Way \cite{Pshirkov:2016qhu}. Since HAWC is sensitive to gamma-rays in the TeV regime, we are interested in extrapolating the SED beyond 1 TeV. To do so, we fit the data above 50 GeV to a power law, and integrate the best-fit power law (green curve in Figure~\ref{fig:SED}) from 1~TeV to infinity and multiply by the solid angle of the bubbles to obtain the total integral flux above 1 TeV over the entire bubble-region. To extrapolate this flux to what we might expect from a similar sized source at the position of M31, we scale the integral flux by $\left(r_\text{MW} / r_\text{M31}\right)^2$. We find an expected integral flux of $3.9\times10^{-15}$~cm$^{-2}$~s$^{-1}$ at energies $>$~1~TeV.

Likewise, if we consider the constant fit covering 0.3 - 100 GeV and calculate the total integral flux expected from M31 bubbles, we obtain $2.06\times10^{-10}$~cm$^{-2}$~s$^{-1}$. In comparison, \cite{Pshirkov:2016qhu} observes an integral flux of $2.6 \pm 0.6\times10^{-9}$~cm$^{-2}$~s$^{-1}$ with a spectral index of $\Gamma_\text{FB} = 2.3 \pm 0.1$ in this energy range. As Andromeda is around three times the size of the Milky Way, it is reasonable to expect that such a structure around M31 would be larger, and therefore brighter, than one around the Milky Way. Just as how the GeV flux observed in \cite{Pshirkov:2016qhu} is larger than what would be expected from a Milky Way sized object, we expect the TeV flux from M31 to also be larger than this extrapolation.

\section{Model Template}

\begin{figure}[t]
\centering
\includegraphics[width=0.5\textwidth]{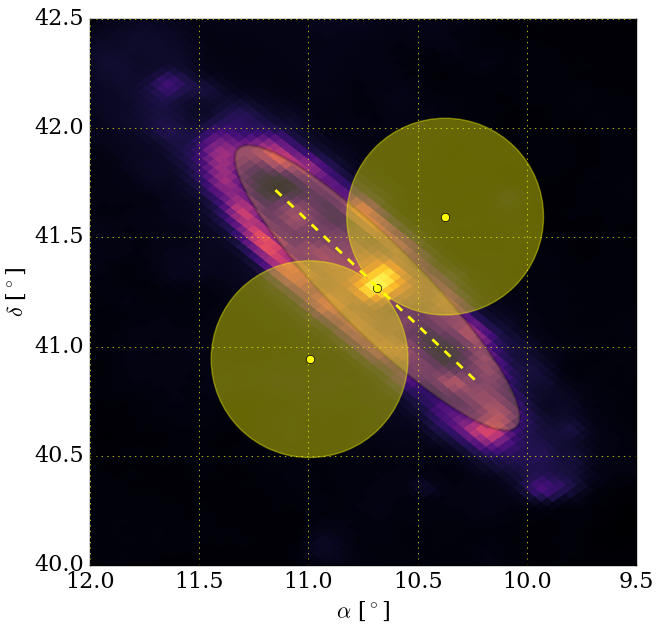}
\caption{Regions to be used in the model for the Fermi Bubbles around M31. The M31 galactic plane is modeled as an ellipse with the two Fermi Bubbles as circular regions with radius $0.45^\circ$. The background image is the IRIS 100$~\mu m$ map of the $2.5^\circ \times 2.5^\circ$ region surrounding M31. The dashed line represents the M31 galactic plane between the two endpoints defined in Table~\ref{tab:positions}.}
\label{fig:bubblemodel}
\end{figure}

We treat the morphology of the M31 Fermi Bubbles as described in \cite{Pshirkov:2016qhu}. The position of the core of M31 is taken from a 100~$\mu$m infrared map from the IRIS (Improved Reprocessing of the IRAS Survey) database \cite{MivilleDeschenes:2004ci}. The positions we determine are given in Table~\ref{tab:positions}. The M31 galactic disk is modeled as an ellipse with semimajor axis defined by two endpoints inclined at 45.04$^{\circ}$ with an aspect ratio $b/a = 0.22$
\cite{Pshirkov:2016qhu}.  The two Fermi Bubble regions are modeled as circular disks of radius $0.45^\circ$, oriented perpendicular to (and symmetrically above and below) the galactic plane of M31. Figure~\ref{fig:bubblemodel} displays the regions used in the spatial
fit.

\begin{table}[ht]
\centering
\begin{tabular*}{0.75\textwidth}{@{\extracolsep{\fill}} c c c }
\hline
\hspace{10pt}Region & $\alpha_{J2000}$ & $\delta_{J2000}$\hspace{10pt} \\
\hline
\hspace{10pt}M31 Nucleus & $10.6848^\circ$ & $41.7166^\circ$\hspace{10pt} \\ [0.5ex]
\hspace{10pt}M31 End 1 & $11.1509^\circ$ & $40.8256^\circ$\hspace{10pt} \\  
\hspace{10pt}M31 End 2 & $10.2145^\circ$ & $40.8256^\circ$\hspace{10pt} \\
\hspace{10pt}FB 1 Center & $10.3746^\circ$ & $41.5951^\circ$\hspace{10pt} \\
\hspace{10pt}FB 2 Center & $10.9949^\circ$ & $40.9430^\circ$\hspace{10pt} \\
\hline
\end{tabular*}
\caption{Positions of M31 regions from the IRIS database. The two endpoints and nucleus of M31 define the galactic plane, which we use to locate two positions perpendicular and symmetric to the center of the plane that are spaced $0.45^\circ$ from the nucleus. `FB' = `Fermi Bubble.'}
\label{tab:positions}
\end{table}


To fit a model of these regions to the HAWC data, we use the Multi-Mission Maximum Likelihood Framework, or 3ML \cite{Vianello:2015wwa}. For both the M31 disk and bubble regions, we assume the gamma-ray emission is defined by a simple power law
spectrum:
\begin{equation}
dN/dE = K\left(\frac{E}{E_0}\right)^{\Gamma}
\label{eq:powerlaw}
\end{equation}
with normalization $K$, pivot energy $E_0$, and spectral index $\Gamma$. We fix
the spectral index of the disk to $\Gamma=-2.5$ and separately fix the spectral
index of the bubble regions to three separate values $(-2.0,\, -2.5,\, -2.75)$. For each of these values, we perform a maximum likelihood analysis to fit the flux normalization from the M31 disk and bubbles. Note that the normalization and index is assumed to be the same for the two bubbles. In all fits, the pivot energy $E_0=1$~TeV is used.

We run the analysis as follows:
\begin{enumerate}
  \item Find the maximum-likelihood estimators for the normalizations, $\hat{K}_\text{disk}$ and $\hat{K}_\text{FB}$.
  \item Use $\hat{K}_\text{disk}$ and $\hat{K}_\text{FB}$ as an input to a Markov-Chain Monte Carlo (MCMC) in order to
  estimate the distribution of the likelihood around the maximum.
  \begin{itemize} \vspace{-5pt}
  	\item We use a uniform prior for both $K_\text{disk}$ and $K_\text{FB}$ in [0, $10^{-11}$~TeV$^{-1}$~cm$^{-2}$~s$^{-1}$]
  \end{itemize}
  \item If no statistically significant flux is observed, calculate a 95\% UL on the flux using the distribution of the likelihood around the maximum (obtained from the MCMC).
\end{enumerate}

\section{HAWC Observations and Results}

\begin{figure}[t]
  \includegraphics[width=0.31\textwidth]{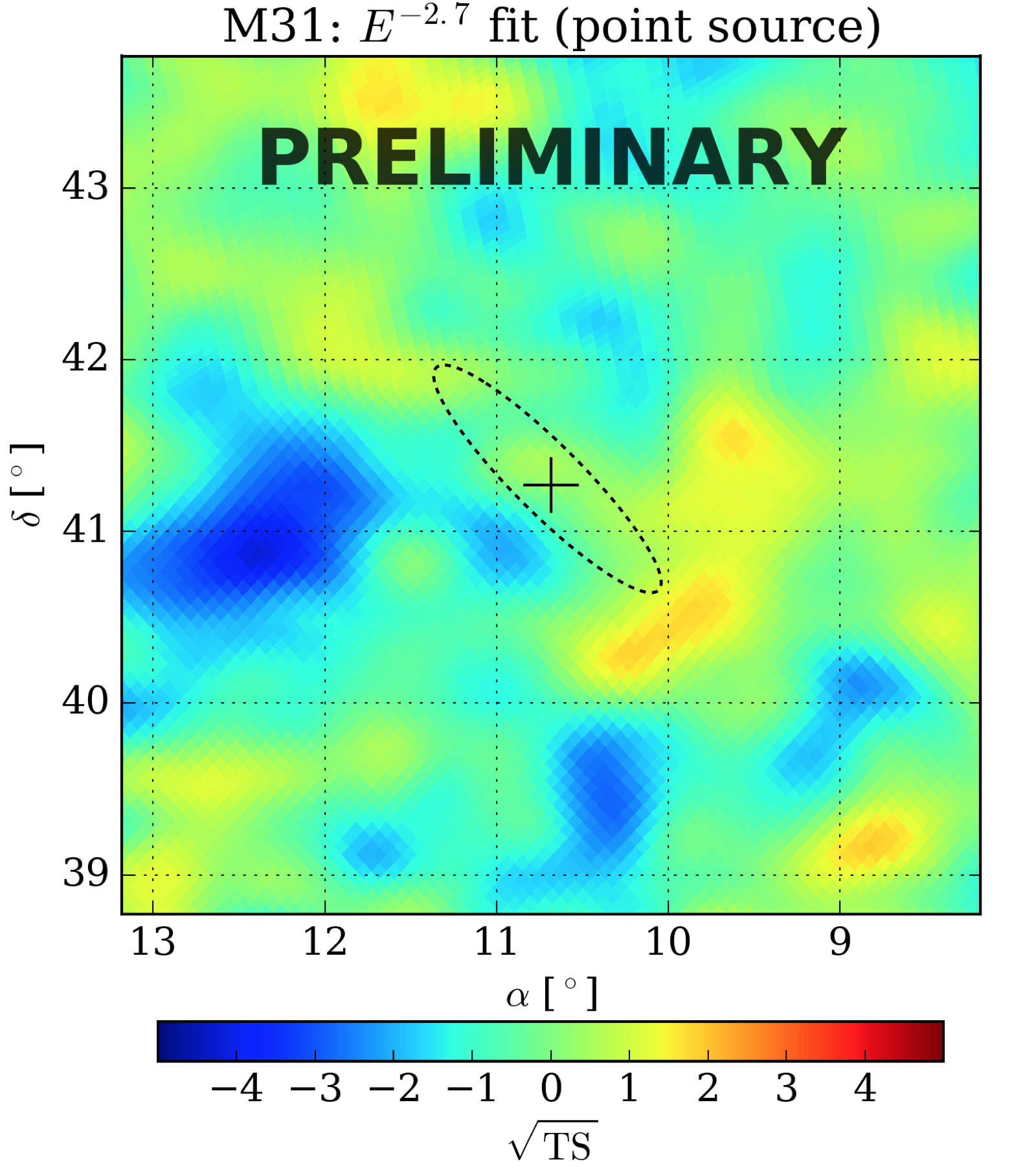}
  \hfill
  \includegraphics[width=0.31\textwidth]{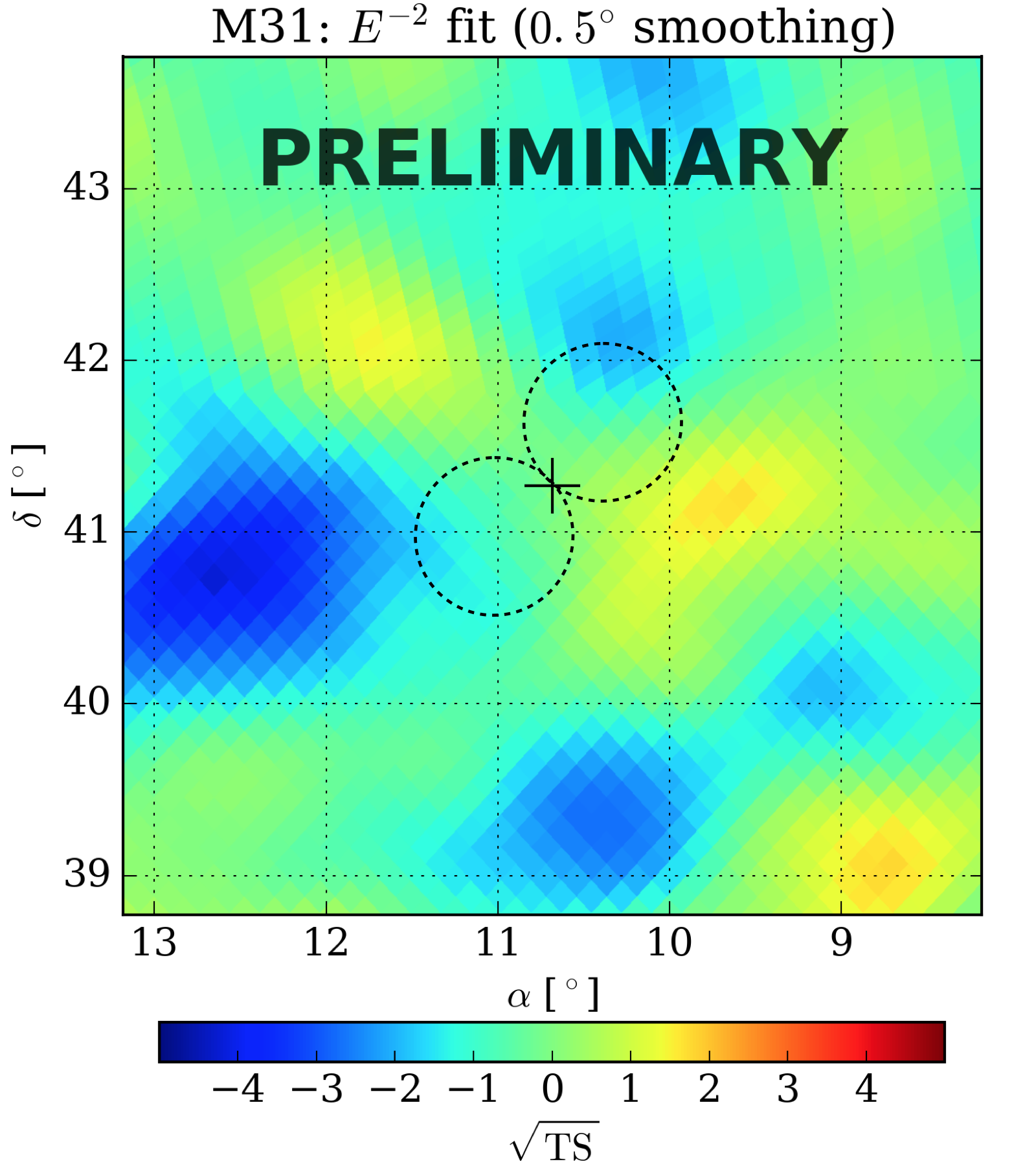}
  \hfill
  \includegraphics[width=0.31\textwidth]{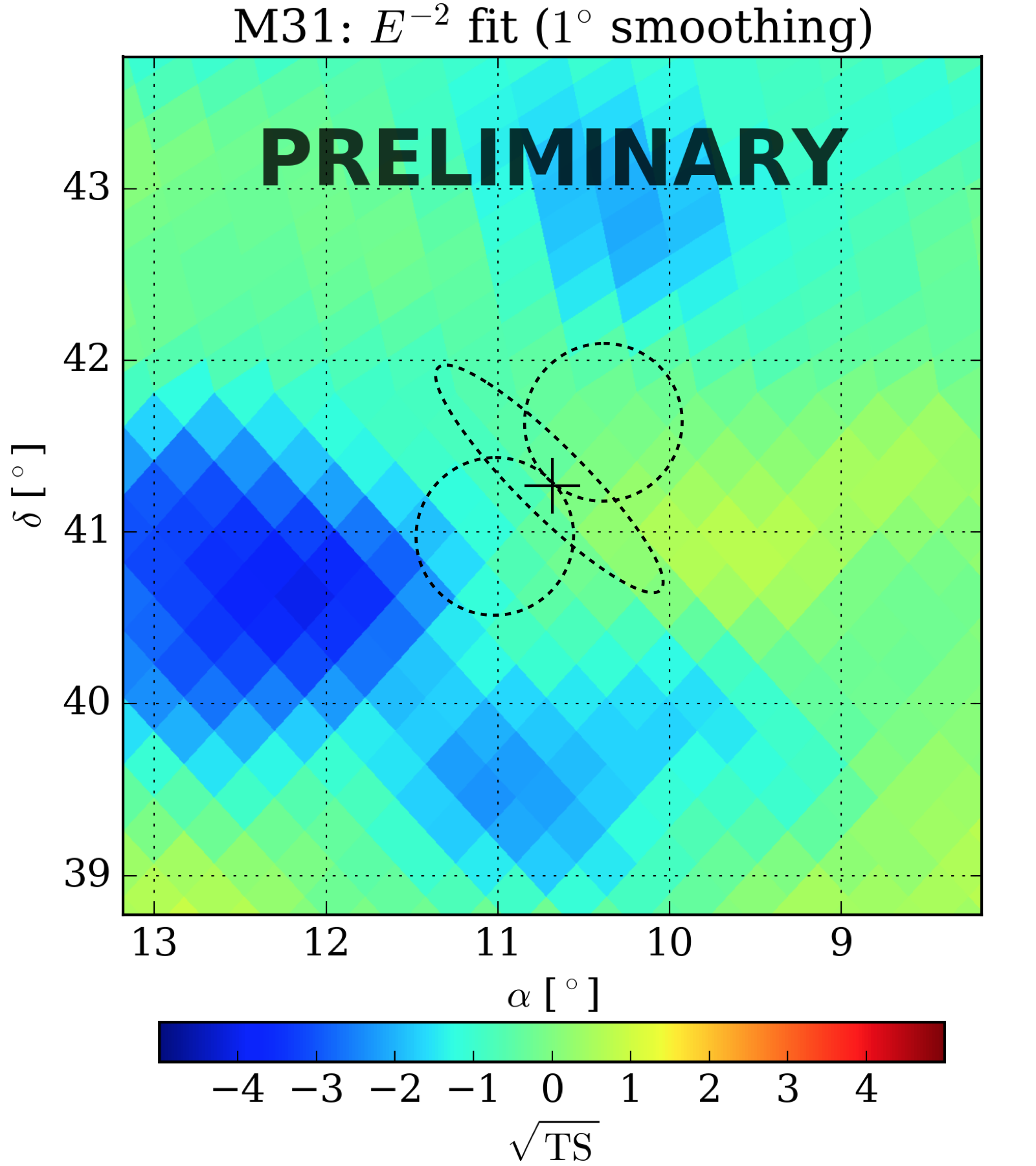}
  \caption{\label{fig:m31_sig_map}
  Significance map in equatorial coordinates of M31 (crosshairs) produced with 25 months of HAWC data
  using the point source map (left), $0.5^\circ$-smoothed map (center), and
  $1^\circ$ extended source map (right). TS (Test Statistic) = $2\log \left(\mathcal{L}_h / \mathcal{L}_0 \right)$, where $\mathcal{L}_0$ is the maximum likelihood given the null hypothesis and $\mathcal{L}_h$ is the maximum likelihood given the alternative hypothesis. $\sqrt{TS} \approx $ Gaussain $\sigma$ (significance). Overlaid on the maps are outlines of the M31 ellipse (left), Fermi Bubbles (center), and combined (right) model regions.}
\end{figure}

We are using 760 days (approximately 25 months) of HAWC data between November 26th 2014 (MJD 56987) and February 18th 2017 (MJD 57802). As can be seen in Figure~\ref{fig:m31_sig_map}, HAWC does not observe any significant excess in gamma ray emission from the region around the location of M31. Thus for each of our model templates, we calculate 95\% upper limits on the flux above 1 TeV. We consider three separate models in this analysis: M31 galactic disk only, two 0.45$^\circ$ disks at the locations of the proposed bubbles, and a combined model including both the disk and two bubbles. From the distribution of the normalizations obtained from the MCMC, a distribution of the integral flux above 1~TeV is calculated, and the 95th percentile is taken as the 95\% upper limit.

We estimate the sensitivity of HAWC to an extended source consisting of an ellipse (M31 galactic disk) and two circular disks (Fermi Bubbles), i.e. our combined model, by fitting such a model to a dataset containing the expected background counts and zero signal events. From this fit we calculate a 95\% upper limit on the integral flux in the same manner described previously, and use this upper limit as the measured sensitivity of HAWC to the M31 region.

\begin{figure}[ht]
\centering
\includegraphics[width=\textwidth]{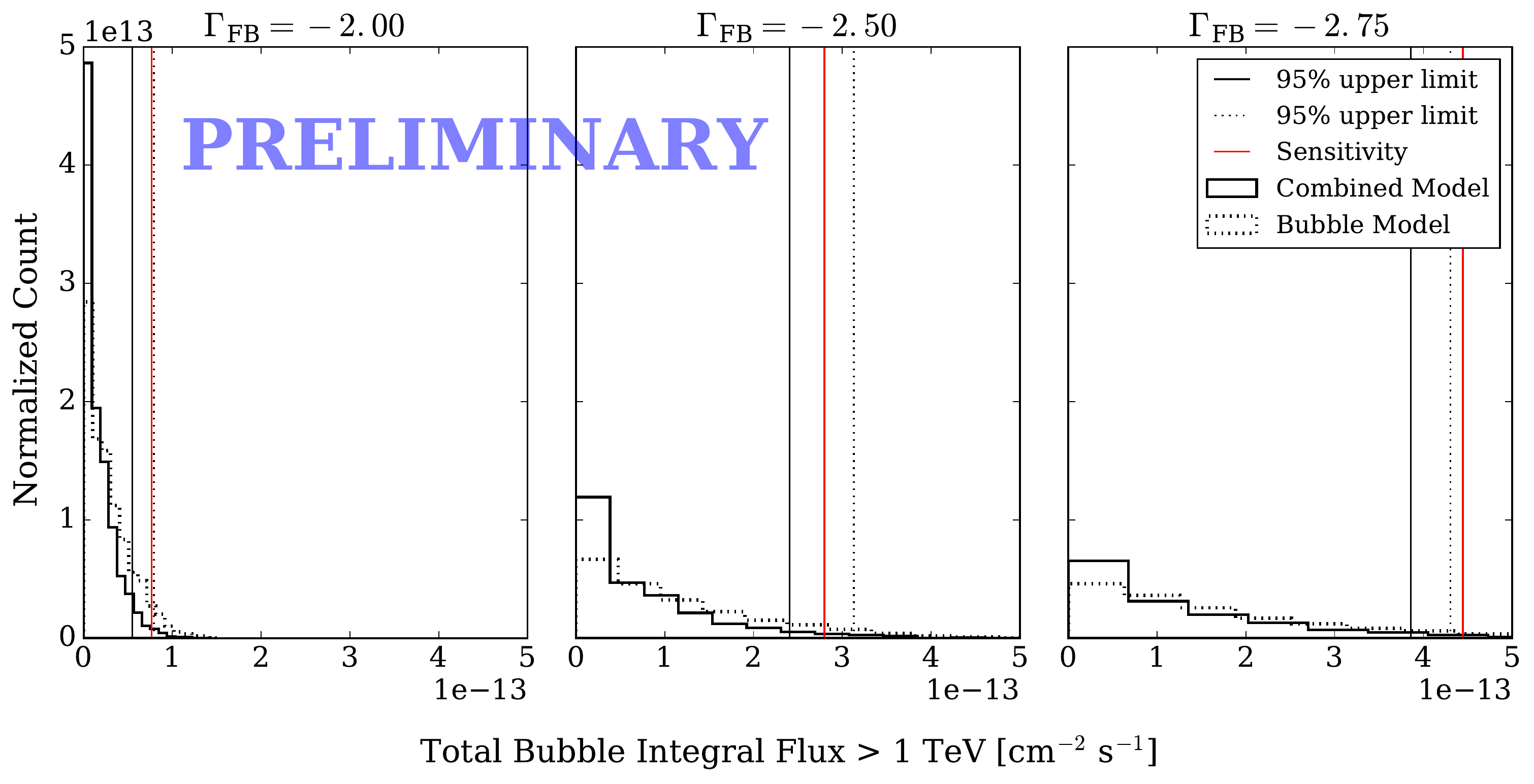}
\caption{MCMC distributions of the total integral flux from two 0.45$^\circ$ bubble regions around M31, for each spectral index tested. The solid line histogram represents the bubble flux when fitting with the combined model (bubbles + M31 disk), and the dotted line represents the bubble flux when fitting with only the two bubble regions (no M31 disk). Vertical lines denote corresponding 95\% upper limits, with the red vertical lines denoting HAWC's sensitivity for each spectral index.}
\label{fig:bubble_flux}
\end{figure}

The distributions of the integral flux from the bubbles for all three spectral indices used in the fit are shown in Figure~\ref{fig:bubble_flux}, along with the measured sensitivity for an extended source consisting of the M31 galactic disk and two circular bubble regions. It can be seen that there is not a very significant change in the distribution or the upper limits when adding the M31 galactic disk to the model on top of the two bubbles, and that the observed limits are approximately the same as the measured sensitivity. The upper limits and sensitivities are tabulated in Table~\ref{tab:bubble_limits} and Table~\ref{tab:sensitivity}, along with the extrapolated integral flux we estimated from the Milky Way bubbles, which can be seen to be well below our upper limits. Upper limits on the flux from the M31 disk can be found in Table~\ref{tab:disk_limits}. 

\begin{table}[ht]
\centering
\begin{tabular*}{0.75\textwidth}{@{\extracolsep{\fill}} c c }
\hline
Spectral Index Used & Integral Flux > 1 TeV  \\
\hline
Sensitivity at $\Gamma_\text{FB} = -2.75$ & $4.44\times10^{-13}\text{ cm}^{-2}\text{ s}^{-1}$ \\
Sensitivity at $\Gamma_\text{FB} = -2.50$ & $2.80\times10^{-13}\text{ cm}^{-2}\text{ s}^{-1}$ \\
Sensitivity at $\Gamma_\text{FB} = -2.00$ & $7.66\times10^{-14}\text{ cm}^{-2}\text{ s}^{-1}$ \\
\hline
\end{tabular*}
\caption{Sensitivity of HAWC to the M31 bubble regions (95\% upper limits on the integral flux from the two bubble regions calculated by fitting the combined model to a background-only dataset).}
\label{tab:sensitivity}
\end{table}

\begin{table}[ht]
\centering
\begin{tabular*}{0.75\textwidth}{@{\extracolsep{\fill}} c c }
\hline
Model / Spectral Index & Integral Flux > 1 TeV  \\
\hline
Extrapolated at $\Gamma_\text{FB} = -2.75$ & $3.93\times10^{-15}\text{ cm}^{-2}\text{ s}^{-1}$ \\ [0.5ex]
\hline
Combined with $\Gamma_\text{FB} = -2.75$ & $3.86\times10^{-13}\text{ cm}^{-2}\text{ s}^{-1}$ \\
Combined with $\Gamma_\text{FB} = -2.50$ & $2.41\times10^{-13}\text{ cm}^{-2}\text{ s}^{-1}$ \\
Combined with $\Gamma_\text{FB} = -2.00$ & $5.50\times10^{-14}\text{ cm}^{-2}\text{ s}^{-1}$ \\
\hline
Bubbles with $\Gamma_\text{FB} = -2.75$ & $4.30\times10^{-13}\text{ cm}^{-2}\text{ s}^{-1}$ \\
Bubbles with $\Gamma_\text{FB} = -2.50$ & $3.13\times10^{-13}\text{ cm}^{-2}\text{ s}^{-1}$ \\
Bubbles with $\Gamma_\text{FB} = -2.00$ & $7.91\times10^{-14}\text{ cm}^{-2}\text{ s}^{-1}$ \\
\hline
\end{tabular*}
\caption{Observed integral flux 95\% upper limits on the emission from the M31 bubble regions.}
\label{tab:bubble_limits}
\end{table}

\clearpage

\begin{table}[h!]
\centering
\begin{tabular*}{0.75\textwidth}{@{\extracolsep{\fill}} c c }
\hline
Model & Integral Flux > 1 TeV  \\
\hline
M31 Disk only & $4.94\times10^{-13}\text{ cm}^{-2}\text{ s}^{-1}$ \\
\hline
Combined with $\Gamma_\text{FB} = -2.75$ & $4.89\times10^{-13}\text{ cm}^{-2}\text{ s}^{-1}$ \\
Combined with $\Gamma_\text{FB} = -2.50$ & $4.45\times10^{-13}\text{ cm}^{-2}\text{ s}^{-1}$ \\
Combined with $\Gamma_\text{FB} = -2.00$ & $4.75\times10^{-13}\text{ cm}^{-2}\text{ s}^{-1}$ \\
\hline
\end{tabular*}
\caption{Observed integral flux 95\% upper limits on the emission from the M31 galactic disk.}
\label{tab:disk_limits}
\end{table}

\section{Conclusion}

We have fit 25 months of HAWC data to spatial templates of the emission of the nearby spiral galaxy Andromeda (M31). Independent and joint fits to the galactic disk and Fermi Bubble-like structures were performed using IRIS templates for the disk emission and bubble templates from a GeV fit performed on Fermi-LAT data.

No significant emission was observed in the region of M31 in the HAWC point source and extended source searches, so we have computed upper limits on the disk and bubble emission. The limits are also compared to the flux detected from the Fermi Bubbles in the Milky Way, normalized to the distance to Andromeda. The upper limits on an M31 Fermi Bubble structure are found to be greater than this extrapolated flux by roughly two orders of magnitude, but are also right at the sensitivity level of HAWC. These upper limits are also roughly four orders of magnitude smaller than the integral flux in the 0.3 - 100 GeV energy range observed using Fermi-LAT data \cite{Pshirkov:2016qhu}. This suggests that the $\gamma$-ray emission from possible Fermi Bubble structures around M31 may exhibit a stronger cutoff than the Milky Way at energies above 100 GeV. This analysis represents the first systematic study of very extended emission from M31 in the TeV range.

%
\bibliographystyle{JHEP}
\bibliography{references}

\end{document}